\documentclass[aps,prl,superscriptaddress,nofootinbib,twocolumn]{revtex4-1}
\usepackage[utf8]{inputenc}
\usepackage{amsmath,amssymb,amsfonts,graphicx,slashed,xspace}
\usepackage[dvipsnames]{xcolor}
\usepackage[colorlinks=true, linkcolor = blue]{hyperref} 
\hypersetup{colorlinks=true, linkcolor=BurntOrange, citecolor=BurntOrange,
filecolor=BurntOrange, urlcolor=BurntOrange}

\newcommand{\bee}{\begin{equation}}
\newcommand{\ene}{\end{equation}}
\newcommand{\bea}{\begin{eqnarray}}
\newcommand{\ena}{\end{eqnarray}}
\def\feynrules{{\sc FeynRules}\xspace}
\def\mgamc{{\sc MG5\_aMC}\xspace}
\def\madanalysis{{\sc MadAnalysis~5}\xspace}
\def\pythia{{\sc Pythia~8}\xspace}
\def\fastjet{{\sc FastJet}\xspace}

\def\fb{\, {\rm fb}}

\def\tev{{\rm TeV}}
\def\gev{{\rm GeV}}

\begin{document}
{\hspace{+15cm} \text{KEK-TH-2304}}\\
%%%%%%%%%%%%%%%%%%%%%%%%%%%%%%%%%%%%%%%%%%%%%

\title{Signatures of toponium formation in LHC run~2 data}

\author{Benjamin Fuks}
\email{fuks@lpthe.jussieu.fr}
\affiliation{Sorbonne Universit\'{e}, CNRS, Laboratoire de Physique Th\'{e}orique et Hautes \'{E}nergies, LPTHE, F-75005 Paris, France}
\affiliation{Institut Universitaire de France, 103 boulevard Saint-Michel, 75005 Paris, France}

\author{Kaoru Hagiwara}
\email{kaoru.hagiwara@kek.jp}
\affiliation{KEK Theory Center, Tsukuba 305-0801, Japan}

\author{Kai Ma}
\email{makai@ucas.ac.cn}
\affiliation{School of Fundamental Physics and Mathematical Science, Hangzhou Institute for Advanced Study, UCAS, Hangzhou 310024, Zhejiang, China}
\affiliation{International Centre for Theoretical Physics Asia-Pacific, Beijing/Hangzhou, China}
\affiliation{Department of Physics, Shaanxi University of Technology, Hanzhong 723000, Shaanxi, China}

\author{Ya-Juan Zheng}
\email{yjzheng@ku.edu}
\affiliation{Department of Physics and Astronomy, University of Kansas, Lawrence, KS 66045, U.S.A.}

\begin{abstract}
We study reported deviations between observations and theoretical predictions associated with the production of a pair of di-leptonically decaying top quarks at the LHC, and we examine the possibility that they reflect a signal of toponium formation. We investigate the production by gluon fusion of a color-singlet spin-0 toponium $\eta_t$ bound state of a top and anti-top quark, that then decays di-leptonically ($gg\to\eta_t \to \bar\ell \ell b \bar{b} \nu \bar{\nu}$). We find strong correlations favoring the production of di-lepton systems featuring a small angular separation in azimuth and a small invariant mass. Although toponium production only contributes to 0.8\% of the total top-quark pair-production cross section at the 13\ \tev\ LHC, there is a possibility that it can account for observed excesses in the narrow edges of phase space. We propose a method to discover toponium formation by `reconstructing' both its top and anti-top quark constituents in the di-lepton channel.
\end{abstract}

\maketitle

%\tableofcontents

%%%%%%%%%%%%%%%%%%%%%%%%%%%%%%%%%%%%%%%%%%%%%%%%%%%%%%%%%%%%%%%%%%%%%%%%%%%%%%%%%%%%%%%%%%%%%%%%%%%%%%%%%%%%%%%%%%%%%%%%%
%
\setcounter{page}{1}
\renewcommand{\thefootnote}{\arabic{footnote}}
\setcounter{footnote}{0}

{\it Introduction} -- Toponium systems consist of bound states of a top and an anti-top quark, and are predicted to impact top pair-production near threshold. Color-singlet toponium ground states should appear either in a $J\!=\!1$ spin triplet state  ($\Theta$) or in a $J\!=\!0$ spin singlet state ($\eta_t$), just like for $J/\psi$ and $\eta_c$ charmonium states and $\Upsilon$ and $\eta_b$ bottomium states. However, contrary to charm or bottom quarkonia, toponium states decay instantly because their top and anti-top constituents are short-lived. Toponium states hence have a large decay width. A correct modeling of their production and decay at colliders should therefore rely on the Green's function of the non-relativistic Hamiltonian that incorporates the effects of both the Coulomb potential and the top quark width~\cite{Fadin:1987wz,Fadin:1988fn}. Using this formalism, we can study the formation of spin-1 toponium at future $e^+e^-$ colliders~\cite{Strassler:1990nw,Sumino:1992ai,Jezabek:1992np}, such as the ILC and/or the FCC-ee, and that of spin-0 toponium produced at hadron colliders in the color-singlet gluon-fusion channel~\cite{Fadin:1990wx}.

The cross section for toponium production at the LHC has been estimated in refs.~\cite{Hagiwara:2008df,Kiyo:2008bv,Sumino:2010bv,Ju:2020otc}, with the corresponding decay properties being considered in ref.~\cite{Sumino:2010bv}. The toponium contributions to the total $t\bar{t}$ production cross section at the 7 and 14 \tev\ LHC have been found to be about 0.90\% and 0.79\% respectively. Because of this relative smallness, the properties of the toponium decay products have not yet been studied carefully.

Recently, the ATLAS collaboration reported a high-statistic measurement of the top pair-production cross section from di-leptonic $t\bar t$ events, and additionally investigated several differential distributions involving both final-state leptons~\cite{Aad:2019hzw}. An excess of events featuring leptons emitted close to each other in azimuth and forming a system with a small invariant mass, {\it i.e.}~with $\Delta\varphi_{\ell\bar\ell}\!\lesssim\! \pi/5$ and $m_{\ell\bar\ell}\!\lesssim\! 20~\gev$, had been observed after a comparison with the best available QCD predictions. These relied on event generation at the next-to-leading-order matched with parton showers (NLO+PS)~\cite{Alioli:2010xd,Alwall:2014hca}, and accounted for a normalization at the next-to-next-to-leading order matched with threshold resummation at the next-to-next-to-leading logarithmic accuracy (NNLO+NNLL)~\cite{Czakon:2013goa}. Whereas this significant excess led to a new physics interpretation through extra scalars~\cite{vonBuddenbrock:2019ajh}, we show in this report that it may be explained as a consequence of spin-0 toponium formation.

\medskip

{\it Angular correlations in di-leptonic $t\bar t$ decays} -- The spin polarization states $M_{\sigma,\bar{\sigma}}$ of the pair of decaying top quarks can be expressed as the helicity amplitudes
\bee
  M_{\sigma,\bar{\sigma}} = M\Big(\eta_t \to t(p,\sigma/2) \ \bar{t}(\bar{p},\bar{\sigma}/2)\Big) \,,
\ene
where $p$ ($\bar{p}$) and $\sigma/2$ ($\bar{\sigma}/2$) are the four-momentum and helicity of the virtual $t$ ($\bar{t}$) quark. As the $s$-wave $\eta_t$ spin-singlet bound state of a top and an anti-top quark is a pseudo-scalar under parity and Lorentz transformations, only the two helicity amplitudes with the same helicities $\sigma\!=\!\bar{\sigma}$ are non-vanishing (the total angular momentum of the system being zero), and they are equal (the $\eta_t$ state being odd under parity), $M_{++} = M_{--}$. The polarization state of the $t\bar{t}$ system is hence determined by the density matrix
\bee
\rho^{\eta_t}_{\sigma\bar{\sigma};\sigma'{\bar\sigma'}}
=
  \frac{M_{\sigma\bar{\sigma}}M_{\sigma'{\bar\sigma'}}^*}{\sum\limits_{\sigma\bar{\sigma}} |M_{\sigma\bar{\sigma}}|^2}
\ene
whose non-vanishing components are all $1/2$ when $\sigma\!=\!\bar{\sigma}\!=\!\pm 1$ and $\sigma'\!=\!\bar{\sigma}'\!=\!\pm 1$. The correlated decay distributions in the di-leptonic modes, with $t \to b W^+ \to b \bar\ell \nu_\ell$ and $\bar{t} \to \bar{b} W^- \to \bar{b} \ell' \bar\nu_{\ell'}$, are then determined from a convolution with the (virtual) top and anti-top decay density matrices $\rho^{t\to b \bar\ell \nu_\ell}_{\sigma,\sigma'}$ and $\rho^{\bar{t}\to \bar{b} \ell' \bar{\nu}_{\ell'}}_{\bar{\sigma},{\bar\sigma'}}$,
\bee\label{eq:densitymat}
\sum_{\sigma,\bar{\sigma},\sigma',{\bar\sigma'}}
\rho^{\eta_t}_{\sigma\bar{\sigma};\sigma'{\bar\sigma'}}\
\rho^{t\to b \bar\ell \nu_\ell}_{\sigma,\sigma'}\
\rho^{\bar{t}\to \bar{b} \ell' \bar{\nu}_{\ell'}}_{\bar{\sigma},{\bar\sigma'}} \,.
\ene
The resulting energy and angular correlations among the top and anti-top decay products near the $t\bar{t}$ production threshold have been studied both for the $e^+e^-\to t\bar{t}$~\cite{Sumino:1992ai,Fujii:1993mk} and $e^+e^-\to Ht\bar{t}$~\cite{Hagiwara:2016rdv,Hagiwara:2017ban,Ma:2018ott} processes, as well as at hadron colliders~\cite{Sumino:2010bv,Ma:2017vve}.

The top and anti-top decay density matrices including the full three-body phase space are, {\it e.g.}, available from ref.~\cite{Hagiwara:2017ban}. If we retain only the dependence on the charged lepton angles in the above distribution, we find
\bee\label{eq:Dis:TopDecay}\begin{split}
& (1+\cos\bar{\theta})(1+\cos\theta)
+ (1-\cos\bar{\theta})(1-\cos\theta)\\
& \qquad\qquad +  2\sin\bar{\theta} \sin\theta \cos(\bar{\varphi}-\varphi)\, .
\end{split}\ene
Here, $\bar{\theta}$ and $\bar{\varphi}$ ($\theta$ and $\varphi$) stand for the polar and azimuthal angle of the charged lepton $\bar\ell$ ($\ell'$) in the top (anti-top) rest frame, defined from the common $z$-axis chosen along the top momentum in the toponium rest frame.

From the form of the correlated decay angular distribution of eq.~\eqref{eq:Dis:TopDecay}, it is clear that the di-lepton pair favors a small invariant mass when the lepton energies are similar. If both top quarks are on-shell and if the toponium mass were exactly at threshold ($m_{\eta_t}\!=\!2m_t$), then the di-lepton invariant mass would be given by
\bee
  m_{\bar{\ell}\ell'}^2 \! =\!  2E_{\bar{\ell}} E_{\ell'} \Big( 1 - \sin\bar{\theta}\sin\theta\cos(\bar{\varphi}-\varphi)- \cos\bar{\theta}\cos\theta \Big)\, .
\ene
The charged lepton energy distribution is flat in the $t$ or $\bar{t}$ rest frame, so that the $m_{\bar{\ell}\ell'}$ invariant mass is small when the differential cross section~\eqref{eq:Dis:TopDecay} is large. Moreover, although the azimuthal angle difference between the two leptons $\Delta\varphi_{\bar{\ell}\ell'}$ measured at the LHC is obtained after boosting the toponium system to the laboratory frame, eq.~\eqref{eq:Dis:TopDecay} still favors small $\Delta\varphi_{\bar{\ell}\ell'}$ values.

\medskip

{\it Toponium production rate at the LHC} -- Encouraged by the above observation, we now examine the toponium production rate at the LHC. Standard event generators used for $t\bar{t}$ measurements at the LHC~\cite{Alioli:2010xd,Alwall:2014hca} provide NLO-accurate predictions that do not include the non-perturbative Coulombic corrections giving rise to toponium formation. We therefore define the toponium production cross section as the difference between the cross section including (${\rm d}\sigma_{\rm full}$) and omitting (${\rm d}\sigma_{\rm NLO}$) these Coulombic corrections,
\bee\label{eq:XS:Definition}
  \sigma(\eta_t)  \equiv
\int_{-8\ \gev}^{4\ \gev} {\rm d} E \left(\frac{{\rm d}\sigma_{\rm full}}{{\rm d}E}
- \frac{{\rm d}\sigma_{\rm NLO}}{{\rm d}E}\right) \, .
\ene
In this expression, we integrate over the toponium binding energy $E \!=\! W-2m_t$ with $W\!=\!m(bW^+\bar{b}W^-)$ being the toponium invariant mass. The differential cross section difference in eq.~\eqref{eq:XS:Definition} exhibits a peak at $E\!\approx\!-2\ \gev$. We therefore fix the integration domain so that $W$ lies in the $[338, 350]$~GeV window around the peak at 344~GeV. This peak value is further loosely identified as the toponium mass $m_{\eta_t}$.

With this definition, we extract the toponium production cross sections $\sigma(\eta_t)$ for colliding energies of $\sqrt{s}=7$ and 14~\tev\ from Figs.~10--11 of ref.~\cite{Sumino:2010bv}, and derive their counterparts at $\sqrt{s}=8$ and 13~TeV by interpolating the results on the basis of the CTEQ6M gluon distribution function~\cite{Pumplin:2002vw}. Our values, listed in table~\ref{tabel:XS:NLO}, are compared with the corresponding NNLO+NNLL $t\bar t$ production cross sections $\sigma(t\bar t)$~\cite{Czakon:2013goa,Czakon:2018nun}, for $m_t\!=\!173.3$~GeV.

%%%%%%%%%%%%%%%%%%%%%%%%%%%%%%%%%%%%%%%%% 
% Table I:
\begin{table}
  \renewcommand\arraystretch{1.0}
  \setlength\tabcolsep{8pt}
  \caption{Toponium and NNLO+NNLL $t\bar{t}$ production cross sections
  in $pp$ collisions at $\sqrt{s}=7$, 8, 13 and 14~TeV. The toponium results at 7 and 14~TeV are obtained from ref.~\cite{Sumino:2010bv}, while those at 8 and 13~\tev\ are derived by interpolation. The $t\bar t$ results are taken from refs.~\cite{Czakon:2013goa,Czakon:2018nun}.\label{tabel:XS:NLO}}
  \begin{center}
    \begin{tabular}{c|c c|c}
      $\sqrt{s}$ & $\sigma(\eta_t)$ [pb] &  $\sigma(t\bar{t})$ [pb] & Ratio \\\hline
      7~\tev  & 1.55 & 172 & 0.0090\\
      8~\tev  & 2.19 & 246 & 0.0089\\
      13~\tev & 6.43 & 810 & 0.0079\\
      14~\tev & 7.54 & 954 & 0.0079
    \end{tabular}
  \end{center}
\end{table}

\medskip

{\it Modeling toponium production at colliders} -- Because toponium production rates are less than $1\%$ of the total $t\bar{t}$ production cross sections (see table~\ref{tabel:XS:NLO}), its impact might have escaped detection so far. In order to assess whether toponium contributions can account for the excess observed by the ATLAS collaboration at low $\Delta\varphi_{\bar\ell\ell'}$ and small $m_{\bar\ell\ell'}$~\cite{Aad:2019hzw}, we should have a suitable event generator. In the absence of any such generator incorporating non-perturbative Coulombic corrections, we propose a toy model built from the effective Lagrangian density
\bee\label{eq:Model:Eta}\begin{split}
  {\cal L}_{\eta_t} =&\ \frac12 \partial_\mu\phi_{\eta_t}\partial^\mu\phi_{\eta_t}
    - \frac12 m_{\eta_t}\phi_{\eta_t}^2\\&\
    - \frac{1}{4}g_{gg\eta_t} \phi_{\eta_t} G^a_{\mu\nu} \tilde G^{a\mu\nu}
    -i g_{tt\eta_t} \phi_{\eta_t} \bar{t} \gamma_5 t.
\end{split}\ene
It involves the pseudo-scalar toponium field $\phi_{\eta_t}(x)$, the covariant gluon field strength tensor $G^a_{\mu\nu}(x)$ (and its dual $\tilde G^a_{\mu\nu}(x)$), the top quark field $t(x)$, and toponium couplings to top quarks ($g_{tt\eta_t}$) and gluons ($g_{gg\eta_t}$).

We import the above effective Lagrangian to \mgamc~\cite{Alwall:2014hca} as a Universal Feynrules Output model~\cite{Degrande:2011ua} generated by \feynrules~\cite{Christensen:2009jx,Alloul:2013bka}. We then generate events in which a toponium state $\eta_t$ is produced in the color-singlet gluon fusion channel, next decays into the six-body final state $b\bar{\ell}\nu_\ell\bar{b}\ell'\bar{\nu}_{\ell'}$ via virtual $t$ and $\bar{t}$ propagators, and where initial-state radiation and hadronization are modeled by \pythia~\cite{Sjostrand:2014zea}. We choose $m_{\eta_t}\!=\!344~\gev$, $\Gamma_{\eta_t}\!=\!7~\gev$ and appropriate couplings to reproduce the results of table~\ref{tabel:XS:NLO}. This value for the $\Gamma_{\eta_t}$ width is obtained by fitting the $m(bW^+\bar{b}W^-)$ distributions of the difference between the QCD predictions with and without Coloumbic corrections from ref.~\cite{Sumino:2010bv}. Details will be reported in ref.~\cite{Fuks:2021vve}.

% $t$ and $\bar{t}$ momentum distribution % 
Although our toy model~\eqref{eq:Model:Eta} with its mass and width fitted to the QCD predictions of ref.~\cite{Sumino:2010bv} accounts for the Coulombic enhancement of the total cross section around the toponium ground state, it does not yield a correct momentum distribution of the $t$ and $\bar{t}$ constituents. The matrix elements given by the Lagrangian \eqref{eq:Model:Eta} are essentially products of $t$ and $\bar{t}$ propagators at a given $W$ value,
\bee
M \sim D_t(s_t)\ D_t(s_{\bar{t}}) \ \ \text{with}\ D_t(s) = \frac{m_t\Gamma_t}{s-m_t^2+im_t\Gamma_t}.
\ene
The Coloumbic corrections however not only affect the overall strengths of the amplitude (which are taken into account as the `resonant' behavior of the $\eta_t$ production amplitude), but also the relative momentum distribution between the $t$ and $\bar{t}$ constituents, which reflects the spatial size of the $\eta_t$ system. The correct momentum distribution at a given $W$ (or $E$) value can be reproduced by enforcing a re-weighting of the squared matrix element,
\bee\label{eq:Reweighting}
  |M|^2 \to |M|^2\ \bigg|\dfrac{G(E; p^*)}{G_0(E; p^*)}\bigg|^2,
\ene
where $p^*$ denotes the common magnitude of the $t$ and $\bar{t}$ momentum in the toponium rest frame, and $G(E; p^*)$ ($G_0(E; p^*)$) denotes the non-relativistic Green’s function of the (free) Hamiltonian.

\medskip

{\it Pinning down toponium formation at the LHC} -- With the above preparation, we generate $t\bar{t}$ events in $pp$ collisions at $\sqrt{s}\!=\!13$~TeV. Predictions without Coulombic corrections rely on standard leading-order event generation matched with parton showers (LO+PS) through \mgamc. We consider the $pp\to W^+bW^-\bar b$ process, followed by leptonic $W$-boson decays, and we normalize the predictions at NNLO+NNLL (see table~\ref{tabel:XS:NLO}). The toponium signal is generated at LO+PS by using the $\eta_t$ model of eq.~\eqref{eq:Model:Eta}, with the same process as for the non-resonant case but with an intermediate $\eta_t$ state. We enforce a generator-level cut of $338~\gev \!<\! W \!<\! 350~\gev$, impose the re-weighting procedure of eq.~\eqref{eq:Reweighting} and normalize the events according to table~\ref{tabel:XS:NLO}. For all simulations, we have used the CT14 set of parton densities~\cite{Hou:2016sho}.

With 140 fb$^{-1}$ of integrated luminosity, we expect about 5,160,000 di-leptonic $t\bar t$ events (with $\ell, \ell'=e,\mu$), whereas 41,000 additional events are predicted for the signal. In order to test whether those events originating from toponium formation below the $t\bar{t}$ threshold can account for the excess of di-lepton events at small $\Delta\varphi_{\bar{\ell}\ell'}$ and small $m_{\bar{\ell}\ell'}$ observed by the ATLAS collaboration~\cite{Aad:2019hzw}, we reconstruct all hadron-level events according to the anti-$k_T$ algorithm~\cite{Cacciari:2008gp} with a radius parameter $R\!=\!0.4$, as implemented in \fastjet~\cite{Cacciari:2011ma}. Next, we implement various selection cuts with \madanalysis~\cite{Conte:2012fm,Conte:2014zja,Conte:2018vmg}, relying on an ideal detector~\cite{Araz:2020lnp} and mimicking the selection of the considered ATLAS analysis.

%%%%%%%%%%%%%%%%%%%%%%%%%%%%%%%%%%%%%%%%% 
% Table II:
\begin{table}
  \renewcommand\arraystretch{1.0}
  \setlength\tabcolsep{6pt}
\caption{Number of expected events at $\sqrt{s}=13~\tev$ for an integrated luminosity of $140~\fb^{-1}$. Predictions for $t\bar t$ production without Coulombic correction and the enhancement due to toponium formation are shown separately, together with their ratio.}
\begin{center}
\begin{tabular}{c|r r|c}
  Cut                             & $t\bar t~~~~~$ & Toponium  & Ratio\\\hline
  Initial                         & 113,000,000 & 900,000 & 0.0079\\
  Di-lepton                       &   5,160,000 &  41,000 & 0.0079\\
  $p_T, |\eta|, \Delta R$         &   1,370,000 &  10,300 & 0.0075\\
  $\Delta\varphi_{\bar\ell\ell'}$ &     178,000 &   4,060 & 0.023\\
  $m_{\bar\ell\ell'}$             &     77,000 &   2,760 & 0.036\\
  $m_T(\bar\ell\ell' b\bar b; \nu_\ell\bar\nu_{\ell'})$ & 40,800 & 2,460 & 0.060\\
  $t\bar t$ kinematical fit       &      20,400 &   1,420 & 0.070
\end{tabular}
\end{center}
\label{tab:cuts}
\end{table}
%%%%%%

We select events including exactly two $b$-jets and two leptons, with a transverse momentum $p_T\!>\!25$~GeV and a pseudo-rapidity $|\eta|\!<\!2.5$. These reconstructed objects are moreover required to be all separated from each other, in the transverse plane, by a distance $\Delta R\!>\!0.4$. About 33.5\% of non-resonant and 25.1\% of toponium events survive the selection, leaving 1,370,000 and 10,300 events respectively, as given in Table~\ref{tab:cuts}.

\begin{figure}\centering
  \includegraphics[width=0.75\columnwidth]{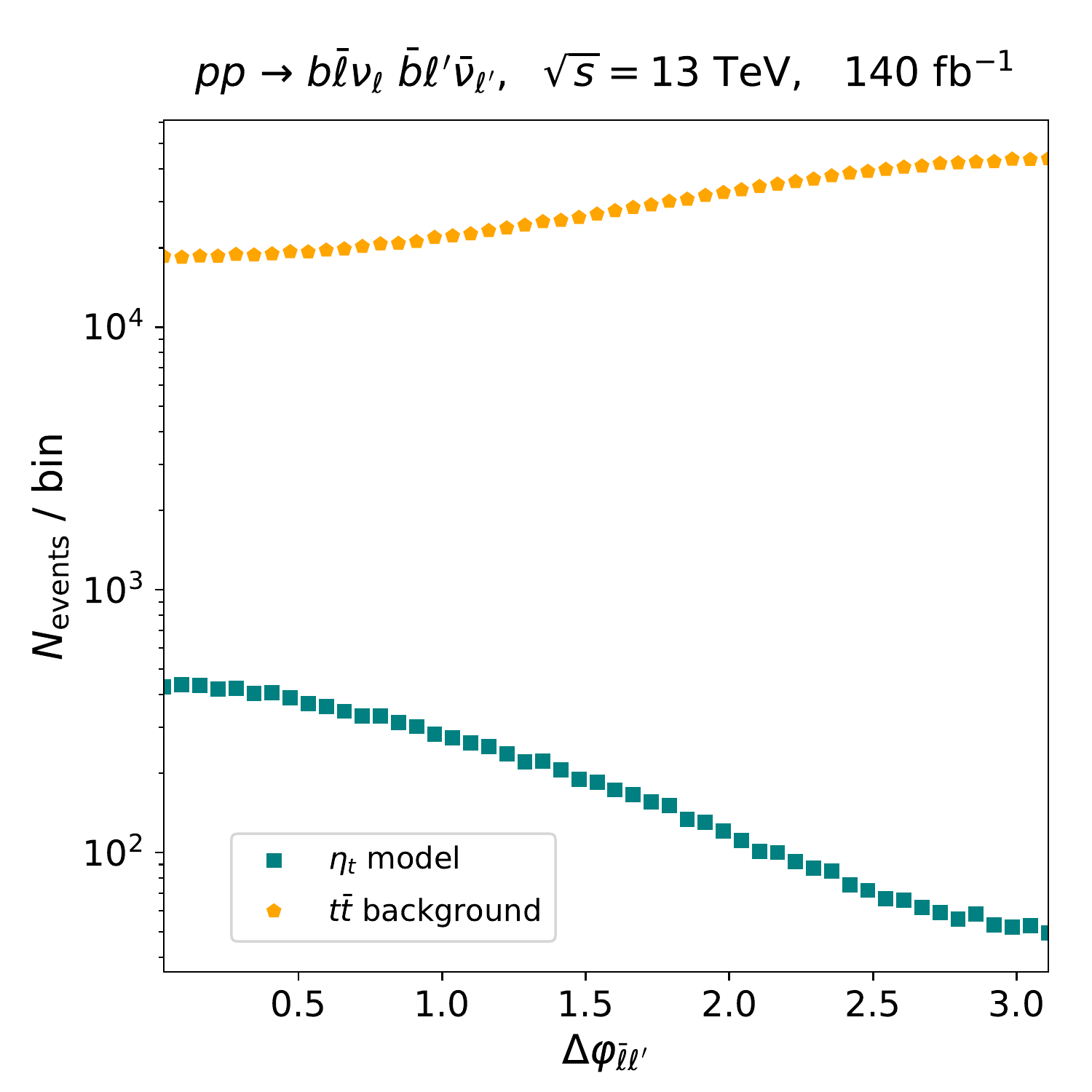}
  \includegraphics[width=0.75\columnwidth]{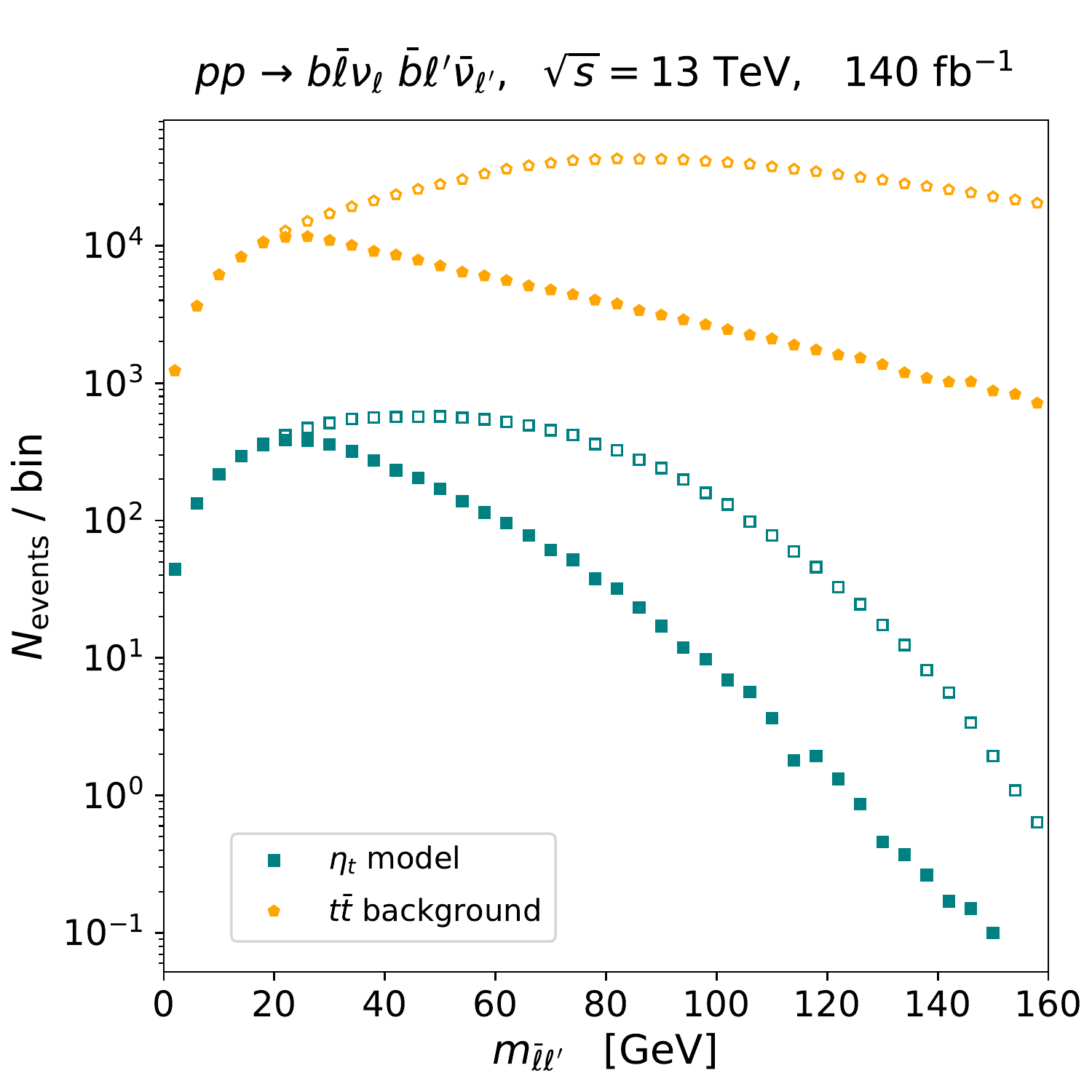}
  \caption{Azimuthal angle difference $\Delta\varphi_{\bar{\ell}\ell'}$ (upper) and invariant mass $m_{\bar\ell\ell'}$ (lower) spectra resulting from di-leptonic events in $t\bar{t}$ production at $13~\tev$. We show predictions for standard $t\bar{t}$ production (orange pentagons) and for toponium formation within the $\eta_t$ model (teal squares) at LO+PS and with normalizations given in table~\ref{tabel:XS:NLO}. The invariant mass distributions are shown before (unfilled markers) and after (filled markers) the $\Delta\varphi_{\bar{\ell}\ell'}\!<\!\pi/5$ cut.}
  \label{fig:Results:AziDiff}
\end{figure}

In the top panel of Fig.~\ref{fig:Results:AziDiff}, we show the $\Delta\varphi_{\bar{\ell}\ell'}$ distribution associated with the toponium signal (teal squares) and that of the non-resonant $t\bar{t}$ background (orange pentagons). We find that toponium contributions strongly prefer small $\Delta\varphi_{\bar{\ell}\ell'}$ values. We hence impose the selection
\bee\label{eq:dphi}
  \Delta\varphi_{\bar{\ell}\ell'} \!<\! \pi/5
\ene
in order to enhance the signal over background ratio. 178,000 non-resonant and 4,060 toponium events survive the cut (see table~\ref{tab:cuts}). In the lower panel of Fig.~\ref{fig:Results:AziDiff}, we present the $m_{\bar{\ell}\ell'}$ distribution before (unfilled markers) and after (filled markers) the cut~\eqref{eq:dphi}. Toponium events favor small $m_{\bar{\ell}\ell'}$ values, and there is a strong positive correlation of events featuring both small $\Delta\varphi_{\bar{\ell}\ell'}$ and $m_{\bar{\ell}\ell'}$ values. This correlation could be regarded as an evidence for toponium formation in ATLAS data, at least if the magnitude of the observed enhancement is consistent with our predictions. Consequently, we require
\bee m_{\bar\ell\ell'}\!<\!40~\gev\, . \ene
In the following, we aim to find a clear signal of toponium formation among the 77,000 non-resonant and 2,760 toponium events that survive this cut, as the signal-over-noise ratio ($S/N$) is still of 3.6\%.

We begin with an additional cut on the transverse mass of the system made of the two $b$-jets, the two leptons and the missing transverse momentum,
\bee
  m_T(\bar\ell\ell' b\bar b; \nu_\ell\bar\nu_{\ell'}) \!<\! 320~\gev\, .
\ene
As almost all toponium events satisfy this cut, the $S/N$ ratio increases to 6.0\%. We continue with by trying out a kinematical reconstruction of the $t$ and $\bar{t}$ toponium constituents from the four-momenta of the reconstructed leptons and $b$-jets, as well as from the missing transverse momentum $\slashed{p}_T$ ({\it i.e.}~the vector sum of the $\nu_\ell$ and $\bar{\nu}_{\ell'}$ transverse momenta). The basic idea behind this reconstruction is that for toponium events, we can make a very rough approximation that the $t$ and $\bar{t}$ transverse momenta are the same. With this assumption, we can reconstruct the $t$ and $\bar{t}$ quarks from a combination of one lepton and one $b$-jet, just like in $t\bar{t}$ events where only one of the quarks decays semi-leptonically. The only remaining complexity is that there is a two-fold ambiguity in distinguishing the two $b$-jets. We resolve it by using the property of the six-body distributions~\eqref{eq:densitymat}.

In practice the two leptons are labeled as $\ell_1$ and $\ell_2$ such that $p_T(\ell_1)>p_T(\ell_2)$, and we define the $b_1$ and $b_2$ jets so that $m(\ell_1,b_1) > m(\ell_1,b_2)$. Next, we assume pair-wise decays, $t_k \to l_k b_k \nu_k$ with $k\!=\!1,2$. The transverse momenta of the neutrinos can be derived from our assumption ${\bf p}_T(t_1)\!=\!{\bf p}_T(t_2)$, and we fix their full three-momenta by assuming that $m(\ell_k, \nu_k)\!=\!m_W$ with a two-fold ambiguity for each neutrino. We choose the solutions leading to $m(b_k,\ell_k,\nu_k)$ invariant masses that are the nearest to the top mass $m_t$. At the end of the reconstruction process, we label the heavier of the reconstructed mass as $m(t_H)$, and the lighter one as $m(t_L)$.

\begin{figure}
  \begin{center}
  \includegraphics[width=0.80\columnwidth]{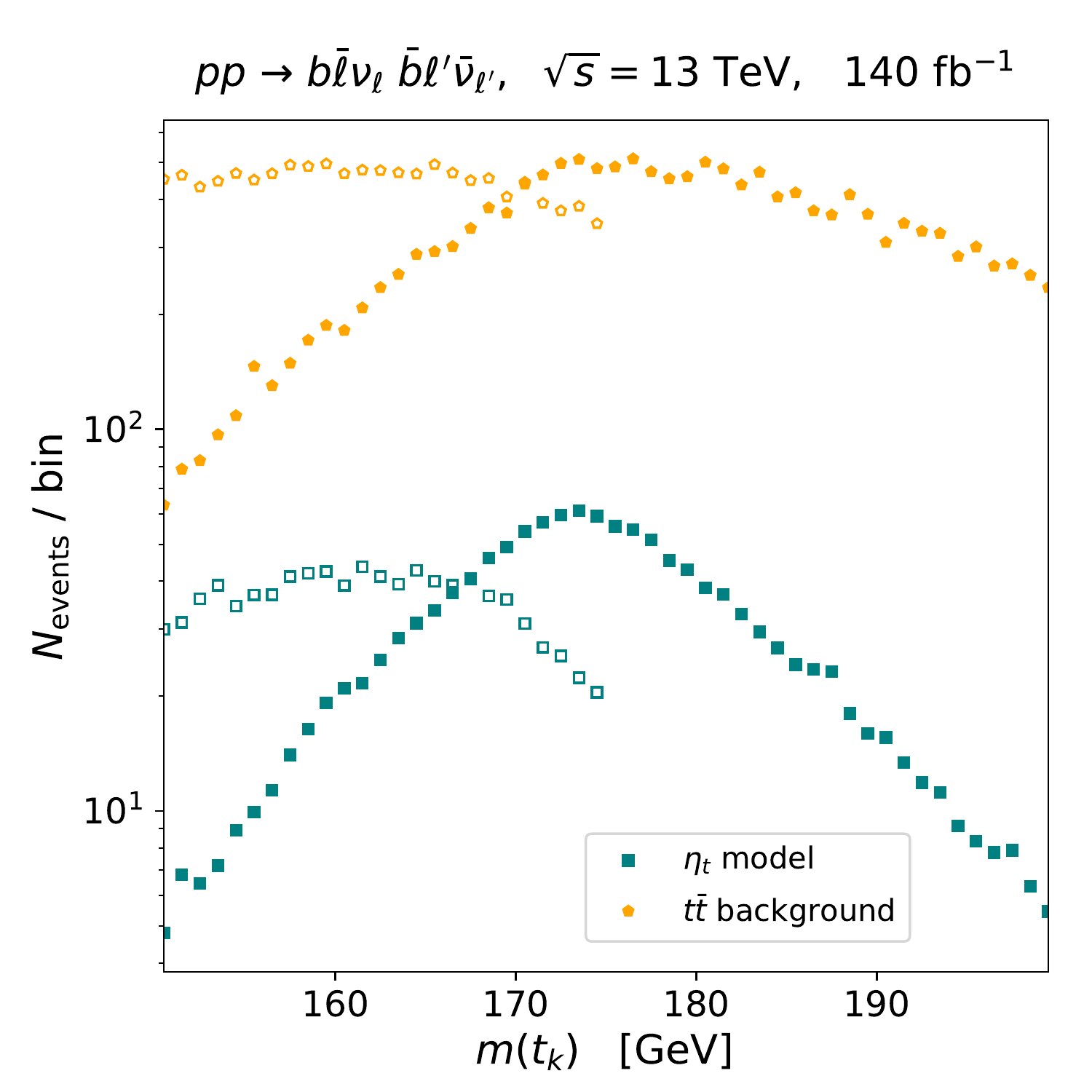}
  \includegraphics[width=0.80\columnwidth]{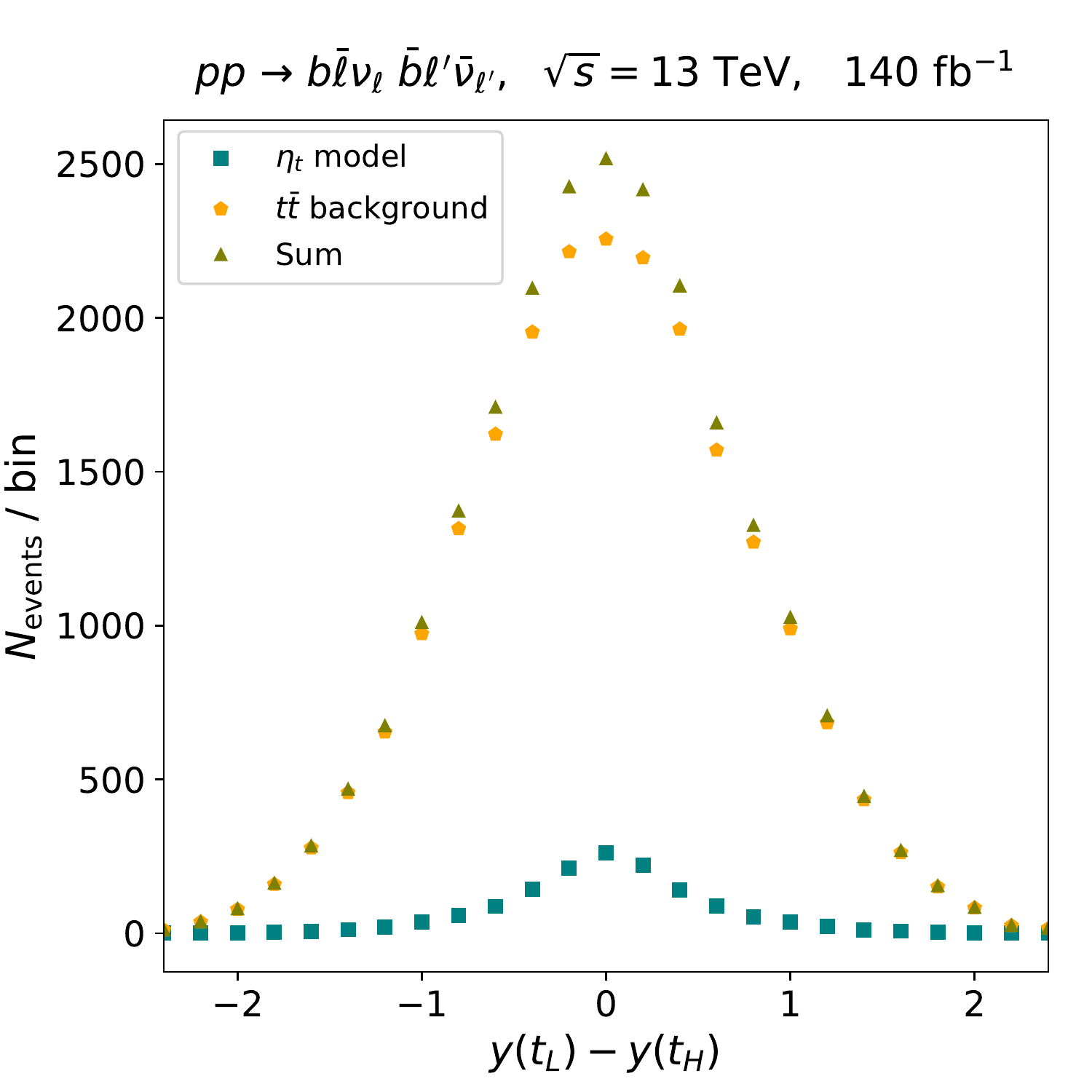}
  \caption{Distributions in the reconstructed top quark invariant masses (upper) and in their rapidity difference (lower). The $m(t_H)$ and $m(t_L)$ spectra are represented with filled and unfilled markers respectively, and we compare predictions for the toponium signal (teal squares) and the non-resonant $t\bar t$ background (orange pentagons). For the rapidity difference spectrum, we also show the sum of the toponium and non-resonant contributions (olive triangles).} \label{fig:reco}
\end{center}
\end{figure}

As shown in table~\ref{tab:cuts}, a significant fraction of the toponium and non-resonant $t\bar{t}$ events can be reconstructed with the above method, leaving 1,420 toponium and 20,400 non-resonant events.
We present in Fig.~\ref{fig:reco} the resulted $m(t_L)$ and $m(t_H)$ distributions, together with the rapidity difference between the toponium constituents $y(t_L)-y(t_H)$. While the $m(t_H)$ toponium distribution exhibits a wide peak centered on $m_t$ (contrary to the background), the reconstructed $t_L$ quark is very offshell and leads to a flat $m(t_L)$ spectrum extending to small values both for the signal and the non-resonant background. On the other hand, the rapidity difference spectrum has a more pronounced peak around the origin for the signal, which should reflect a smaller $p^*$ momentum inside the $\eta_t$ system than is the case for the continuum background. 
Those observables therefore offer interesting handles to characterize toponium formation at the LHC~\cite{Fuks:2021vve}.

\medskip

{\it Conclusions} -- Although our study is still primitive, we hope that dedicated investigations of di-lepton $t\bar t$ events should reveal an evidence for toponium formation along the lines presented in this work. Needless to say, the toponium signal should also be observed in those events where one of the top quark decays hadronically, and the other one semi-leptonically. In this case, the spin-0 toponium decay correlation exhibited in the distribution~\eqref{eq:densitymat} can be identified by using the hadronic decay density matrix of ref.~\cite{Hagiwara:2017ban}, where di-jet systems originating from $W$-boson decays are given half and half weights to stem from a down-type and an up-type quark. In addition, the pseudo-scalar nature of the toponium could be confirmed by studying jet angular correlation in $t\bar{t}$ events featuring two additional jets, as proposed in ref.~\cite{Klamke:2007cu,Hagiwara:2013jp,Nakamura:2015uca}.

%%%%%%%%%%%%%%%%%%%%%%%%%%%%%%%%%%
\section*{Acknowledgements}
We are grateful for stimulating discussions with Hiroshi Yokoya and Yukinari Sumino. BF and KH thank Alan Cornell, Deepak Kar, James Keaveney and Bruce Mellado for explaining the data in great detail. KH thanks Biswarup Mukhopadhaya at HRI and K.C. Kong at University of Kansas, where the study was initiated, and the US Japan Cooperation Program in High Energy Physics for financial support. KM is supported by the National Natural Science Foundation of China under Grant no.~11705113, Natural Science Basic Research Plan in Shaanxi Province of China under Grant no.~2018JQ1018, and the Scientific Research Program Funded by Shaanxi Provincial Education Department under Grant no.~18JK0153. YZ is supported by the US Department of Energy under Grant no.~DE-SC009474.

\bibliography{TopPheno}

\begin{thebibliography}{37}%
\makeatletter
\providecommand \@ifxundefined [1]{%
 \@ifx{#1\undefined}
}%
\providecommand \@ifnum [1]{%
 \ifnum #1\expandafter \@firstoftwo
 \else \expandafter \@secondoftwo
 \fi
}%
\providecommand \@ifx [1]{%
 \ifx #1\expandafter \@firstoftwo
 \else \expandafter \@secondoftwo
 \fi
}%
\providecommand \natexlab [1]{#1}%
\providecommand \enquote  [1]{``#1''}%
\providecommand \bibnamefont  [1]{#1}%
\providecommand \bibfnamefont [1]{#1}%
\providecommand \citenamefont [1]{#1}%
\providecommand \href@noop [0]{\@secondoftwo}%
\providecommand \href [0]{\begingroup \@sanitize@url \@href}%
\providecommand \@href[1]{\@@startlink{#1}\@@href}%
\providecommand \@@href[1]{\endgroup#1\@@endlink}%
\providecommand \@sanitize@url [0]{\catcode `\\12\catcode `\$12\catcode
  `\&12\catcode `\#12\catcode `\^12\catcode `\_12\catcode `\%12\relax}%
\providecommand \@@startlink[1]{}%
\providecommand \@@endlink[0]{}%
\providecommand \url  [0]{\begingroup\@sanitize@url \@url }%
\providecommand \@url [1]{\endgroup\@href {#1}{\urlprefix }}%
\providecommand \urlprefix  [0]{URL }%
\providecommand \Eprint [0]{\href }%
\providecommand \doibase [0]{http://dx.doi.org/}%
\providecommand \selectlanguage [0]{\@gobble}%
\providecommand \bibinfo  [0]{\@secondoftwo}%
\providecommand \bibfield  [0]{\@secondoftwo}%
\providecommand \translation [1]{[#1]}%
\providecommand \BibitemOpen [0]{}%
\providecommand \bibitemStop [0]{}%
\providecommand \bibitemNoStop [0]{.\EOS\space}%
\providecommand \EOS [0]{\spacefactor3000\relax}%
\providecommand \BibitemShut  [1]{\csname bibitem#1\endcsname}%
\let\auto@bib@innerbib\@empty
%</preamble>
\bibitem [{\citenamefont {Fadin}\ and\ \citenamefont
  {Khoze}(1987)}]{Fadin:1987wz}%
  \BibitemOpen
  \bibfield  {author} {\bibinfo {author} {\bibfnamefont {V.~S.}\ \bibnamefont
  {Fadin}}\ and\ \bibinfo {author} {\bibfnamefont {V.~A.}\ \bibnamefont
  {Khoze}},\ }\href@noop {} {\bibfield  {journal} {\bibinfo  {journal} {JETP
  Lett.}\ }\textbf {\bibinfo {volume} {46}},\ \bibinfo {pages} {525} (\bibinfo
  {year} {1987})}\BibitemShut {NoStop}%
\bibitem [{\citenamefont {Fadin}\ and\ \citenamefont
  {Khoze}(1988)}]{Fadin:1988fn}%
  \BibitemOpen
  \bibfield  {author} {\bibinfo {author} {\bibfnamefont {V.~S.}\ \bibnamefont
  {Fadin}}\ and\ \bibinfo {author} {\bibfnamefont {V.~A.}\ \bibnamefont
  {Khoze}},\ }\href@noop {} {\bibfield  {journal} {\bibinfo  {journal} {Sov. J.
  Nucl. Phys.}\ }\textbf {\bibinfo {volume} {48}},\ \bibinfo {pages} {309}
  (\bibinfo {year} {1988})}\BibitemShut {NoStop}%
\bibitem [{\citenamefont {Strassler}\ and\ \citenamefont
  {Peskin}(1991)}]{Strassler:1990nw}%
  \BibitemOpen
  \bibfield  {author} {\bibinfo {author} {\bibfnamefont {M.~J.}\ \bibnamefont
  {Strassler}}\ and\ \bibinfo {author} {\bibfnamefont {M.~E.}\ \bibnamefont
  {Peskin}},\ }\href {\doibase 10.1103/PhysRevD.43.1500} {\bibfield  {journal}
  {\bibinfo  {journal} {Phys. Rev. D}\ }\textbf {\bibinfo {volume} {43}},\
  \bibinfo {pages} {1500} (\bibinfo {year} {1991})}\BibitemShut {NoStop}%
\bibitem [{\citenamefont {Sumino}\ \emph {et~al.}(1993)\citenamefont {Sumino},
  \citenamefont {Fujii}, \citenamefont {Hagiwara}, \citenamefont {Murayama},\
  and\ \citenamefont {Ng}}]{Sumino:1992ai}%
  \BibitemOpen
  \bibfield  {author} {\bibinfo {author} {\bibfnamefont {Y.}~\bibnamefont
  {Sumino}}, \bibinfo {author} {\bibfnamefont {K.}~\bibnamefont {Fujii}},
  \bibinfo {author} {\bibfnamefont {K.}~\bibnamefont {Hagiwara}}, \bibinfo
  {author} {\bibfnamefont {H.}~\bibnamefont {Murayama}}, \ and\ \bibinfo
  {author} {\bibfnamefont {C.~K.}\ \bibnamefont {Ng}},\ }\href {\doibase
  10.1103/PhysRevD.47.56} {\bibfield  {journal} {\bibinfo  {journal} {Phys.
  Rev. D}\ }\textbf {\bibinfo {volume} {47}},\ \bibinfo {pages} {56} (\bibinfo
  {year} {1993})}\BibitemShut {NoStop}%
\bibitem [{\citenamefont {Jezabek}\ \emph {et~al.}(1992)\citenamefont
  {Jezabek}, \citenamefont {Kuhn},\ and\ \citenamefont
  {Teubner}}]{Jezabek:1992np}%
  \BibitemOpen
  \bibfield  {author} {\bibinfo {author} {\bibfnamefont {M.}~\bibnamefont
  {Jezabek}}, \bibinfo {author} {\bibfnamefont {J.~H.}\ \bibnamefont {Kuhn}}, \
  and\ \bibinfo {author} {\bibfnamefont {T.}~\bibnamefont {Teubner}},\ }\href
  {\doibase 10.1007/BF01474740} {\bibfield  {journal} {\bibinfo  {journal} {Z.
  Phys. C}\ }\textbf {\bibinfo {volume} {56}},\ \bibinfo {pages} {653}
  (\bibinfo {year} {1992})}\BibitemShut {NoStop}%
\bibitem [{\citenamefont {Fadin}\ \emph {et~al.}(1990)\citenamefont {Fadin},
  \citenamefont {Khoze},\ and\ \citenamefont {Sjostrand}}]{Fadin:1990wx}%
  \BibitemOpen
  \bibfield  {author} {\bibinfo {author} {\bibfnamefont {V.~S.}\ \bibnamefont
  {Fadin}}, \bibinfo {author} {\bibfnamefont {V.~A.}\ \bibnamefont {Khoze}}, \
  and\ \bibinfo {author} {\bibfnamefont {T.}~\bibnamefont {Sjostrand}},\ }\href
  {\doibase 10.1007/BF01614696} {\bibfield  {journal} {\bibinfo  {journal} {Z.
  Phys. C}\ }\textbf {\bibinfo {volume} {48}},\ \bibinfo {pages} {613}
  (\bibinfo {year} {1990})}\BibitemShut {NoStop}%
\bibitem [{\citenamefont {Hagiwara}\ \emph {et~al.}(2008)\citenamefont
  {Hagiwara}, \citenamefont {Sumino},\ and\ \citenamefont
  {Yokoya}}]{Hagiwara:2008df}%
  \BibitemOpen
  \bibfield  {author} {\bibinfo {author} {\bibfnamefont {K.}~\bibnamefont
  {Hagiwara}}, \bibinfo {author} {\bibfnamefont {Y.}~\bibnamefont {Sumino}}, \
  and\ \bibinfo {author} {\bibfnamefont {H.}~\bibnamefont {Yokoya}},\ }\href
  {\doibase 10.1016/j.physletb.2008.07.006} {\bibfield  {journal} {\bibinfo
  {journal} {Phys. Lett. B}\ }\textbf {\bibinfo {volume} {666}},\ \bibinfo
  {pages} {71} (\bibinfo {year} {2008})},\ \Eprint
  {http://arxiv.org/abs/0804.1014} {arXiv:0804.1014 [hep-ph]} \BibitemShut
  {NoStop}%
\bibitem [{\citenamefont {Kiyo}\ \emph {et~al.}(2009)\citenamefont {Kiyo},
  \citenamefont {Kuhn}, \citenamefont {Moch}, \citenamefont {Steinhauser},\
  and\ \citenamefont {Uwer}}]{Kiyo:2008bv}%
  \BibitemOpen
  \bibfield  {author} {\bibinfo {author} {\bibfnamefont {Y.}~\bibnamefont
  {Kiyo}}, \bibinfo {author} {\bibfnamefont {J.~H.}\ \bibnamefont {Kuhn}},
  \bibinfo {author} {\bibfnamefont {S.}~\bibnamefont {Moch}}, \bibinfo {author}
  {\bibfnamefont {M.}~\bibnamefont {Steinhauser}}, \ and\ \bibinfo {author}
  {\bibfnamefont {P.}~\bibnamefont {Uwer}},\ }\href {\doibase
  10.1140/epjc/s10052-009-0892-7} {\bibfield  {journal} {\bibinfo  {journal}
  {Eur. Phys. J. C}\ }\textbf {\bibinfo {volume} {60}},\ \bibinfo {pages} {375}
  (\bibinfo {year} {2009})},\ \Eprint {http://arxiv.org/abs/0812.0919}
  {arXiv:0812.0919 [hep-ph]} \BibitemShut {NoStop}%
\bibitem [{\citenamefont {Sumino}\ and\ \citenamefont
  {Yokoya}(2010)}]{Sumino:2010bv}%
  \BibitemOpen
  \bibfield  {author} {\bibinfo {author} {\bibfnamefont {Y.}~\bibnamefont
  {Sumino}}\ and\ \bibinfo {author} {\bibfnamefont {H.}~\bibnamefont
  {Yokoya}},\ }\href {\doibase 10.1007/JHEP09(2010)034} {\bibfield  {journal}
  {\bibinfo  {journal} {JHEP}\ }\textbf {\bibinfo {volume} {09}},\ \bibinfo
  {pages} {034} (\bibinfo {year} {2010})},\ \bibinfo {note} {[Erratum: JHEP 06,
  037 (2016)]},\ \Eprint {http://arxiv.org/abs/1007.0075} {arXiv:1007.0075
  [hep-ph]} \BibitemShut {NoStop}%
\bibitem [{\citenamefont {Ju}\ \emph {et~al.}(2020)\citenamefont {Ju},
  \citenamefont {Wang}, \citenamefont {Wang}, \citenamefont {Xu}, \citenamefont
  {Xu},\ and\ \citenamefont {Yang}}]{Ju:2020otc}%
  \BibitemOpen
  \bibfield  {author} {\bibinfo {author} {\bibfnamefont {W.-L.}\ \bibnamefont
  {Ju}}, \bibinfo {author} {\bibfnamefont {G.}~\bibnamefont {Wang}}, \bibinfo
  {author} {\bibfnamefont {X.}~\bibnamefont {Wang}}, \bibinfo {author}
  {\bibfnamefont {X.}~\bibnamefont {Xu}}, \bibinfo {author} {\bibfnamefont
  {Y.}~\bibnamefont {Xu}}, \ and\ \bibinfo {author} {\bibfnamefont {L.~L.}\
  \bibnamefont {Yang}},\ }\href {\doibase 10.1007/JHEP06(2020)158} {\bibfield
  {journal} {\bibinfo  {journal} {JHEP}\ }\textbf {\bibinfo {volume} {06}},\
  \bibinfo {pages} {158} (\bibinfo {year} {2020})},\ \Eprint
  {http://arxiv.org/abs/2004.03088} {arXiv:2004.03088 [hep-ph]} \BibitemShut
  {NoStop}%
\bibitem [{\citenamefont {Aad}\ \emph {et~al.}(2020)\citenamefont {Aad} \emph
  {et~al.}}]{Aad:2019hzw}%
  \BibitemOpen
  \bibfield  {author} {\bibinfo {author} {\bibfnamefont {G.}~\bibnamefont
  {Aad}} \emph {et~al.} (\bibinfo {collaboration} {ATLAS}),\ }\href {\doibase
  10.1140/epjc/s10052-020-7907-9} {\bibfield  {journal} {\bibinfo  {journal}
  {Eur. Phys. J. C}\ }\textbf {\bibinfo {volume} {80}},\ \bibinfo {pages} {528}
  (\bibinfo {year} {2020})},\ \Eprint {http://arxiv.org/abs/1910.08819}
  {arXiv:1910.08819 [hep-ex]} \BibitemShut {NoStop}%
\bibitem [{\citenamefont {Alioli}\ \emph {et~al.}(2010)\citenamefont {Alioli},
  \citenamefont {Nason}, \citenamefont {Oleari},\ and\ \citenamefont
  {Re}}]{Alioli:2010xd}%
  \BibitemOpen
  \bibfield  {author} {\bibinfo {author} {\bibfnamefont {S.}~\bibnamefont
  {Alioli}}, \bibinfo {author} {\bibfnamefont {P.}~\bibnamefont {Nason}},
  \bibinfo {author} {\bibfnamefont {C.}~\bibnamefont {Oleari}}, \ and\ \bibinfo
  {author} {\bibfnamefont {E.}~\bibnamefont {Re}},\ }\href {\doibase
  10.1007/JHEP06(2010)043} {\bibfield  {journal} {\bibinfo  {journal} {JHEP}\
  }\textbf {\bibinfo {volume} {06}},\ \bibinfo {pages} {043} (\bibinfo {year}
  {2010})},\ \Eprint {http://arxiv.org/abs/1002.2581} {arXiv:1002.2581
  [hep-ph]} \BibitemShut {NoStop}%
\bibitem [{\citenamefont {Alwall}\ \emph {et~al.}(2014)\citenamefont {Alwall},
  \citenamefont {Frederix}, \citenamefont {Frixione}, \citenamefont {Hirschi},
  \citenamefont {Maltoni}, \citenamefont {Mattelaer}, \citenamefont {Shao},
  \citenamefont {Stelzer}, \citenamefont {Torrielli},\ and\ \citenamefont
  {Zaro}}]{Alwall:2014hca}%
  \BibitemOpen
  \bibfield  {author} {\bibinfo {author} {\bibfnamefont {J.}~\bibnamefont
  {Alwall}}, \bibinfo {author} {\bibfnamefont {R.}~\bibnamefont {Frederix}},
  \bibinfo {author} {\bibfnamefont {S.}~\bibnamefont {Frixione}}, \bibinfo
  {author} {\bibfnamefont {V.}~\bibnamefont {Hirschi}}, \bibinfo {author}
  {\bibfnamefont {F.}~\bibnamefont {Maltoni}}, \bibinfo {author} {\bibfnamefont
  {O.}~\bibnamefont {Mattelaer}}, \bibinfo {author} {\bibfnamefont {H.~S.}\
  \bibnamefont {Shao}}, \bibinfo {author} {\bibfnamefont {T.}~\bibnamefont
  {Stelzer}}, \bibinfo {author} {\bibfnamefont {P.}~\bibnamefont {Torrielli}},
  \ and\ \bibinfo {author} {\bibfnamefont {M.}~\bibnamefont {Zaro}},\ }\href
  {\doibase 10.1007/JHEP07(2014)079} {\bibfield  {journal} {\bibinfo  {journal}
  {JHEP}\ }\textbf {\bibinfo {volume} {07}},\ \bibinfo {pages} {079} (\bibinfo
  {year} {2014})},\ \Eprint {http://arxiv.org/abs/1405.0301} {arXiv:1405.0301
  [hep-ph]} \BibitemShut {NoStop}%
\bibitem [{\citenamefont {Czakon}\ \emph {et~al.}(2013)\citenamefont {Czakon},
  \citenamefont {Fiedler},\ and\ \citenamefont {Mitov}}]{Czakon:2013goa}%
  \BibitemOpen
  \bibfield  {author} {\bibinfo {author} {\bibfnamefont {M.}~\bibnamefont
  {Czakon}}, \bibinfo {author} {\bibfnamefont {P.}~\bibnamefont {Fiedler}}, \
  and\ \bibinfo {author} {\bibfnamefont {A.}~\bibnamefont {Mitov}},\ }\href
  {\doibase 10.1103/PhysRevLett.110.252004} {\bibfield  {journal} {\bibinfo
  {journal} {Phys. Rev. Lett.}\ }\textbf {\bibinfo {volume} {110}},\ \bibinfo
  {pages} {252004} (\bibinfo {year} {2013})},\ \Eprint
  {http://arxiv.org/abs/1303.6254} {arXiv:1303.6254 [hep-ph]} \BibitemShut
  {NoStop}%
\bibitem [{\citenamefont {Buddenbrock}\ \emph {et~al.}(2019)\citenamefont
  {Buddenbrock}, \citenamefont {Cornell}, \citenamefont {Fang}, \citenamefont
  {Fadol~Mohammed}, \citenamefont {Kumar}, \citenamefont {Mellado},\ and\
  \citenamefont {Tomiwa}}]{vonBuddenbrock:2019ajh}%
  \BibitemOpen
  \bibfield  {author} {\bibinfo {author} {\bibfnamefont {S.}~\bibnamefont
  {Buddenbrock}}, \bibinfo {author} {\bibfnamefont {A.~S.}\ \bibnamefont
  {Cornell}}, \bibinfo {author} {\bibfnamefont {Y.}~\bibnamefont {Fang}},
  \bibinfo {author} {\bibfnamefont {A.}~\bibnamefont {Fadol~Mohammed}},
  \bibinfo {author} {\bibfnamefont {M.}~\bibnamefont {Kumar}}, \bibinfo
  {author} {\bibfnamefont {B.}~\bibnamefont {Mellado}}, \ and\ \bibinfo
  {author} {\bibfnamefont {K.~G.}\ \bibnamefont {Tomiwa}},\ }\href {\doibase
  10.1007/JHEP10(2019)157} {\bibfield  {journal} {\bibinfo  {journal} {JHEP}\
  }\textbf {\bibinfo {volume} {10}},\ \bibinfo {pages} {157} (\bibinfo {year}
  {2019})},\ \Eprint {http://arxiv.org/abs/1901.05300} {arXiv:1901.05300
  [hep-ph]} \BibitemShut {NoStop}%
\bibitem [{\citenamefont {Fujii}\ \emph {et~al.}(1994)\citenamefont {Fujii},
  \citenamefont {Matsui},\ and\ \citenamefont {Sumino}}]{Fujii:1993mk}%
  \BibitemOpen
  \bibfield  {author} {\bibinfo {author} {\bibfnamefont {K.}~\bibnamefont
  {Fujii}}, \bibinfo {author} {\bibfnamefont {T.}~\bibnamefont {Matsui}}, \
  and\ \bibinfo {author} {\bibfnamefont {Y.}~\bibnamefont {Sumino}},\ }\href
  {\doibase 10.1103/PhysRevD.50.4341} {\bibfield  {journal} {\bibinfo
  {journal} {Phys. Rev. D}\ }\textbf {\bibinfo {volume} {50}},\ \bibinfo
  {pages} {4341} (\bibinfo {year} {1994})}\BibitemShut {NoStop}%
\bibitem [{\citenamefont {Hagiwara}\ \emph {et~al.}(2016)\citenamefont
  {Hagiwara}, \citenamefont {Ma},\ and\ \citenamefont
  {Yokoya}}]{Hagiwara:2016rdv}%
  \BibitemOpen
  \bibfield  {author} {\bibinfo {author} {\bibfnamefont {K.}~\bibnamefont
  {Hagiwara}}, \bibinfo {author} {\bibfnamefont {K.}~\bibnamefont {Ma}}, \ and\
  \bibinfo {author} {\bibfnamefont {H.}~\bibnamefont {Yokoya}},\ }\href
  {\doibase 10.1007/JHEP06(2016)048} {\bibfield  {journal} {\bibinfo  {journal}
  {JHEP}\ }\textbf {\bibinfo {volume} {06}},\ \bibinfo {pages} {048} (\bibinfo
  {year} {2016})},\ \Eprint {http://arxiv.org/abs/1602.00684} {arXiv:1602.00684
  [hep-ph]} \BibitemShut {NoStop}%
\bibitem [{\citenamefont {Hagiwara}\ \emph {et~al.}(2018)\citenamefont
  {Hagiwara}, \citenamefont {Yokoya},\ and\ \citenamefont
  {Zheng}}]{Hagiwara:2017ban}%
  \BibitemOpen
  \bibfield  {author} {\bibinfo {author} {\bibfnamefont {K.}~\bibnamefont
  {Hagiwara}}, \bibinfo {author} {\bibfnamefont {H.}~\bibnamefont {Yokoya}}, \
  and\ \bibinfo {author} {\bibfnamefont {Y.-J.}\ \bibnamefont {Zheng}},\ }\href
  {\doibase 10.1007/JHEP02(2018)180} {\bibfield  {journal} {\bibinfo  {journal}
  {JHEP}\ }\textbf {\bibinfo {volume} {02}},\ \bibinfo {pages} {180} (\bibinfo
  {year} {2018})},\ \Eprint {http://arxiv.org/abs/1712.09953} {arXiv:1712.09953
  [hep-ph]} \BibitemShut {NoStop}%
\bibitem [{\citenamefont {Ma}(2019)}]{Ma:2018ott}%
  \BibitemOpen
  \bibfield  {author} {\bibinfo {author} {\bibfnamefont {K.}~\bibnamefont
  {Ma}},\ }\href {\doibase 10.1016/j.physletb.2019.134928} {\bibfield
  {journal} {\bibinfo  {journal} {Phys. Lett. B}\ }\textbf {\bibinfo {volume}
  {797}},\ \bibinfo {pages} {134928} (\bibinfo {year} {2019})},\ \Eprint
  {http://arxiv.org/abs/1809.07127} {arXiv:1809.07127 [hep-ph]} \BibitemShut
  {NoStop}%
\bibitem [{\citenamefont {Ma}(2017)}]{Ma:2017vve}%
  \BibitemOpen
  \bibfield  {author} {\bibinfo {author} {\bibfnamefont {K.}~\bibnamefont
  {Ma}},\ }\href {\doibase 10.1103/PhysRevD.96.071501} {\bibfield  {journal}
  {\bibinfo  {journal} {Phys. Rev. D}\ }\textbf {\bibinfo {volume} {96}},\
  \bibinfo {pages} {071501} (\bibinfo {year} {2017})},\ \Eprint
  {http://arxiv.org/abs/1706.01492} {arXiv:1706.01492 [hep-ph]} \BibitemShut
  {NoStop}%
\bibitem [{\citenamefont {Pumplin}\ \emph {et~al.}(2002)\citenamefont
  {Pumplin}, \citenamefont {Stump}, \citenamefont {Huston}, \citenamefont
  {Lai}, \citenamefont {Nadolsky},\ and\ \citenamefont
  {Tung}}]{Pumplin:2002vw}%
  \BibitemOpen
  \bibfield  {author} {\bibinfo {author} {\bibfnamefont {J.}~\bibnamefont
  {Pumplin}}, \bibinfo {author} {\bibfnamefont {D.~R.}\ \bibnamefont {Stump}},
  \bibinfo {author} {\bibfnamefont {J.}~\bibnamefont {Huston}}, \bibinfo
  {author} {\bibfnamefont {H.~L.}\ \bibnamefont {Lai}}, \bibinfo {author}
  {\bibfnamefont {P.~M.}\ \bibnamefont {Nadolsky}}, \ and\ \bibinfo {author}
  {\bibfnamefont {W.~K.}\ \bibnamefont {Tung}},\ }\href {\doibase
  10.1088/1126-6708/2002/07/012} {\bibfield  {journal} {\bibinfo  {journal}
  {JHEP}\ }\textbf {\bibinfo {volume} {07}},\ \bibinfo {pages} {012} (\bibinfo
  {year} {2002})},\ \Eprint {http://arxiv.org/abs/hep-ph/0201195}
  {arXiv:hep-ph/0201195} \BibitemShut {NoStop}%
\bibitem [{\citenamefont {Czakon}\ \emph {et~al.}(2018)\citenamefont {Czakon},
  \citenamefont {Ferroglia}, \citenamefont {Heymes}, \citenamefont {Mitov},
  \citenamefont {Pecjak}, \citenamefont {Scott}, \citenamefont {Wang},\ and\
  \citenamefont {Yang}}]{Czakon:2018nun}%
  \BibitemOpen
  \bibfield  {author} {\bibinfo {author} {\bibfnamefont {M.}~\bibnamefont
  {Czakon}}, \bibinfo {author} {\bibfnamefont {A.}~\bibnamefont {Ferroglia}},
  \bibinfo {author} {\bibfnamefont {D.}~\bibnamefont {Heymes}}, \bibinfo
  {author} {\bibfnamefont {A.}~\bibnamefont {Mitov}}, \bibinfo {author}
  {\bibfnamefont {B.~D.}\ \bibnamefont {Pecjak}}, \bibinfo {author}
  {\bibfnamefont {D.~J.}\ \bibnamefont {Scott}}, \bibinfo {author}
  {\bibfnamefont {X.}~\bibnamefont {Wang}}, \ and\ \bibinfo {author}
  {\bibfnamefont {L.~L.}\ \bibnamefont {Yang}},\ }\href {\doibase
  10.1007/JHEP05(2018)149} {\bibfield  {journal} {\bibinfo  {journal} {JHEP}\
  }\textbf {\bibinfo {volume} {05}},\ \bibinfo {pages} {149} (\bibinfo {year}
  {2018})},\ \Eprint {http://arxiv.org/abs/1803.07623} {arXiv:1803.07623
  [hep-ph]} \BibitemShut {NoStop}%
\bibitem [{\citenamefont {Degrande}\ \emph {et~al.}(2012)\citenamefont
  {Degrande}, \citenamefont {Duhr}, \citenamefont {Fuks}, \citenamefont
  {Grellscheid}, \citenamefont {Mattelaer},\ and\ \citenamefont
  {Reiter}}]{Degrande:2011ua}%
  \BibitemOpen
  \bibfield  {author} {\bibinfo {author} {\bibfnamefont {C.}~\bibnamefont
  {Degrande}}, \bibinfo {author} {\bibfnamefont {C.}~\bibnamefont {Duhr}},
  \bibinfo {author} {\bibfnamefont {B.}~\bibnamefont {Fuks}}, \bibinfo {author}
  {\bibfnamefont {D.}~\bibnamefont {Grellscheid}}, \bibinfo {author}
  {\bibfnamefont {O.}~\bibnamefont {Mattelaer}}, \ and\ \bibinfo {author}
  {\bibfnamefont {T.}~\bibnamefont {Reiter}},\ }\href {\doibase
  10.1016/j.cpc.2012.01.022} {\bibfield  {journal} {\bibinfo  {journal}
  {Comput. Phys. Commun.}\ }\textbf {\bibinfo {volume} {183}},\ \bibinfo
  {pages} {1201} (\bibinfo {year} {2012})},\ \Eprint
  {http://arxiv.org/abs/1108.2040} {arXiv:1108.2040 [hep-ph]} \BibitemShut
  {NoStop}%
\bibitem [{\citenamefont {Christensen}\ \emph {et~al.}(2011)\citenamefont
  {Christensen}, \citenamefont {de~Aquino}, \citenamefont {Degrande},
  \citenamefont {Duhr}, \citenamefont {Fuks}, \citenamefont {Herquet},
  \citenamefont {Maltoni},\ and\ \citenamefont
  {Schumann}}]{Christensen:2009jx}%
  \BibitemOpen
  \bibfield  {author} {\bibinfo {author} {\bibfnamefont {N.~D.}\ \bibnamefont
  {Christensen}}, \bibinfo {author} {\bibfnamefont {P.}~\bibnamefont
  {de~Aquino}}, \bibinfo {author} {\bibfnamefont {C.}~\bibnamefont {Degrande}},
  \bibinfo {author} {\bibfnamefont {C.}~\bibnamefont {Duhr}}, \bibinfo {author}
  {\bibfnamefont {B.}~\bibnamefont {Fuks}}, \bibinfo {author} {\bibfnamefont
  {M.}~\bibnamefont {Herquet}}, \bibinfo {author} {\bibfnamefont
  {F.}~\bibnamefont {Maltoni}}, \ and\ \bibinfo {author} {\bibfnamefont
  {S.}~\bibnamefont {Schumann}},\ }\href {\doibase
  10.1140/epjc/s10052-011-1541-5} {\bibfield  {journal} {\bibinfo  {journal}
  {Eur. Phys. J. C}\ }\textbf {\bibinfo {volume} {71}},\ \bibinfo {pages}
  {1541} (\bibinfo {year} {2011})},\ \Eprint {http://arxiv.org/abs/0906.2474}
  {arXiv:0906.2474 [hep-ph]} \BibitemShut {NoStop}%
\bibitem [{\citenamefont {Alloul}\ \emph {et~al.}(2014)\citenamefont {Alloul},
  \citenamefont {Christensen}, \citenamefont {Degrande}, \citenamefont {Duhr},\
  and\ \citenamefont {Fuks}}]{Alloul:2013bka}%
  \BibitemOpen
  \bibfield  {author} {\bibinfo {author} {\bibfnamefont {A.}~\bibnamefont
  {Alloul}}, \bibinfo {author} {\bibfnamefont {N.~D.}\ \bibnamefont
  {Christensen}}, \bibinfo {author} {\bibfnamefont {C.}~\bibnamefont
  {Degrande}}, \bibinfo {author} {\bibfnamefont {C.}~\bibnamefont {Duhr}}, \
  and\ \bibinfo {author} {\bibfnamefont {B.}~\bibnamefont {Fuks}},\ }\href
  {\doibase 10.1016/j.cpc.2014.04.012} {\bibfield  {journal} {\bibinfo
  {journal} {Comput. Phys. Commun.}\ }\textbf {\bibinfo {volume} {185}},\
  \bibinfo {pages} {2250} (\bibinfo {year} {2014})},\ \Eprint
  {http://arxiv.org/abs/1310.1921} {arXiv:1310.1921 [hep-ph]} \BibitemShut
  {NoStop}%
\bibitem [{\citenamefont {Sj\"ostrand}\ \emph {et~al.}(2015)\citenamefont
  {Sj\"ostrand}, \citenamefont {Ask}, \citenamefont {Christiansen},
  \citenamefont {Corke}, \citenamefont {Desai}, \citenamefont {Ilten},
  \citenamefont {Mrenna}, \citenamefont {Prestel}, \citenamefont {Rasmussen},\
  and\ \citenamefont {Skands}}]{Sjostrand:2014zea}%
  \BibitemOpen
  \bibfield  {author} {\bibinfo {author} {\bibfnamefont {T.}~\bibnamefont
  {Sj\"ostrand}}, \bibinfo {author} {\bibfnamefont {S.}~\bibnamefont {Ask}},
  \bibinfo {author} {\bibfnamefont {J.~R.}\ \bibnamefont {Christiansen}},
  \bibinfo {author} {\bibfnamefont {R.}~\bibnamefont {Corke}}, \bibinfo
  {author} {\bibfnamefont {N.}~\bibnamefont {Desai}}, \bibinfo {author}
  {\bibfnamefont {P.}~\bibnamefont {Ilten}}, \bibinfo {author} {\bibfnamefont
  {S.}~\bibnamefont {Mrenna}}, \bibinfo {author} {\bibfnamefont
  {S.}~\bibnamefont {Prestel}}, \bibinfo {author} {\bibfnamefont {C.~O.}\
  \bibnamefont {Rasmussen}}, \ and\ \bibinfo {author} {\bibfnamefont {P.~Z.}\
  \bibnamefont {Skands}},\ }\href {\doibase 10.1016/j.cpc.2015.01.024}
  {\bibfield  {journal} {\bibinfo  {journal} {Comput. Phys. Commun.}\ }\textbf
  {\bibinfo {volume} {191}},\ \bibinfo {pages} {159} (\bibinfo {year}
  {2015})},\ \Eprint {http://arxiv.org/abs/1410.3012} {arXiv:1410.3012
  [hep-ph]} \BibitemShut {NoStop}%
\bibitem [{\citenamefont {Fuks}\ \emph {et~al.}()\citenamefont {Fuks},
  \citenamefont {Hagiwara}, \citenamefont {Ma},\ and\ \citenamefont
  {Zheng}}]{Fuks:2021vve}%
  \BibitemOpen
  \bibfield  {author} {\bibinfo {author} {\bibfnamefont {B.}~\bibnamefont
  {Fuks}}, \bibinfo {author} {\bibfnamefont {K.}~\bibnamefont {Hagiwara}},
  \bibinfo {author} {\bibfnamefont {K.}~\bibnamefont {Ma}}, \ and\ \bibinfo
  {author} {\bibfnamefont {Y.-J.}\ \bibnamefont {Zheng}},\ }\href@noop {}
  {\bibinfo  {journal} {{in preparation}}\ }\BibitemShut {NoStop}%
\bibitem [{\citenamefont {Hou}\ \emph {et~al.}(2017)\citenamefont {Hou} \emph
  {et~al.}}]{Hou:2016sho}%
  \BibitemOpen
\bibfield  {journal} {  }\bibfield  {author} {\bibinfo {author} {\bibfnamefont
  {T.-J.}\ \bibnamefont {Hou}} \emph {et~al.},\ }\href {\doibase
  10.1007/JHEP03(2017)099} {\bibfield  {journal} {\bibinfo  {journal} {JHEP}\
  }\textbf {\bibinfo {volume} {03}},\ \bibinfo {pages} {099} (\bibinfo {year}
  {2017})},\ \Eprint {http://arxiv.org/abs/1607.06066} {arXiv:1607.06066
  [hep-ph]} \BibitemShut {NoStop}%
\bibitem [{\citenamefont {Cacciari}\ \emph {et~al.}(2008)\citenamefont
  {Cacciari}, \citenamefont {Salam},\ and\ \citenamefont
  {Soyez}}]{Cacciari:2008gp}%
  \BibitemOpen
  \bibfield  {author} {\bibinfo {author} {\bibfnamefont {M.}~\bibnamefont
  {Cacciari}}, \bibinfo {author} {\bibfnamefont {G.~P.}\ \bibnamefont {Salam}},
  \ and\ \bibinfo {author} {\bibfnamefont {G.}~\bibnamefont {Soyez}},\ }\href
  {\doibase 10.1088/1126-6708/2008/04/063} {\bibfield  {journal} {\bibinfo
  {journal} {JHEP}\ }\textbf {\bibinfo {volume} {04}},\ \bibinfo {pages} {063}
  (\bibinfo {year} {2008})},\ \Eprint {http://arxiv.org/abs/0802.1189}
  {arXiv:0802.1189 [hep-ph]} \BibitemShut {NoStop}%
\bibitem [{\citenamefont {Cacciari}\ \emph {et~al.}(2012)\citenamefont
  {Cacciari}, \citenamefont {Salam},\ and\ \citenamefont
  {Soyez}}]{Cacciari:2011ma}%
  \BibitemOpen
  \bibfield  {author} {\bibinfo {author} {\bibfnamefont {M.}~\bibnamefont
  {Cacciari}}, \bibinfo {author} {\bibfnamefont {G.~P.}\ \bibnamefont {Salam}},
  \ and\ \bibinfo {author} {\bibfnamefont {G.}~\bibnamefont {Soyez}},\ }\href
  {\doibase 10.1140/epjc/s10052-012-1896-2} {\bibfield  {journal} {\bibinfo
  {journal} {Eur. Phys. J. C}\ }\textbf {\bibinfo {volume} {72}},\ \bibinfo
  {pages} {1896} (\bibinfo {year} {2012})},\ \Eprint
  {http://arxiv.org/abs/1111.6097} {arXiv:1111.6097 [hep-ph]} \BibitemShut
  {NoStop}%
\bibitem [{\citenamefont {Conte}\ \emph {et~al.}(2013)\citenamefont {Conte},
  \citenamefont {Fuks},\ and\ \citenamefont {Serret}}]{Conte:2012fm}%
  \BibitemOpen
  \bibfield  {author} {\bibinfo {author} {\bibfnamefont {E.}~\bibnamefont
  {Conte}}, \bibinfo {author} {\bibfnamefont {B.}~\bibnamefont {Fuks}}, \ and\
  \bibinfo {author} {\bibfnamefont {G.}~\bibnamefont {Serret}},\ }\href
  {\doibase 10.1016/j.cpc.2012.09.009} {\bibfield  {journal} {\bibinfo
  {journal} {Comput. Phys. Commun.}\ }\textbf {\bibinfo {volume} {184}},\
  \bibinfo {pages} {222} (\bibinfo {year} {2013})},\ \Eprint
  {http://arxiv.org/abs/1206.1599} {arXiv:1206.1599 [hep-ph]} \BibitemShut
  {NoStop}%
\bibitem [{\citenamefont {Conte}\ \emph {et~al.}(2014)\citenamefont {Conte},
  \citenamefont {Dumont}, \citenamefont {Fuks},\ and\ \citenamefont
  {Wymant}}]{Conte:2014zja}%
  \BibitemOpen
  \bibfield  {author} {\bibinfo {author} {\bibfnamefont {E.}~\bibnamefont
  {Conte}}, \bibinfo {author} {\bibfnamefont {B.}~\bibnamefont {Dumont}},
  \bibinfo {author} {\bibfnamefont {B.}~\bibnamefont {Fuks}}, \ and\ \bibinfo
  {author} {\bibfnamefont {C.}~\bibnamefont {Wymant}},\ }\href {\doibase
  10.1140/epjc/s10052-014-3103-0} {\bibfield  {journal} {\bibinfo  {journal}
  {Eur. Phys. J. C}\ }\textbf {\bibinfo {volume} {74}},\ \bibinfo {pages}
  {3103} (\bibinfo {year} {2014})},\ \Eprint {http://arxiv.org/abs/1405.3982}
  {arXiv:1405.3982 [hep-ph]} \BibitemShut {NoStop}%
\bibitem [{\citenamefont {Conte}\ and\ \citenamefont
  {Fuks}(2018)}]{Conte:2018vmg}%
  \BibitemOpen
  \bibfield  {author} {\bibinfo {author} {\bibfnamefont {E.}~\bibnamefont
  {Conte}}\ and\ \bibinfo {author} {\bibfnamefont {B.}~\bibnamefont {Fuks}},\
  }\href {\doibase 10.1142/S0217751X18300272} {\bibfield  {journal} {\bibinfo
  {journal} {Int. J. Mod. Phys. A}\ }\textbf {\bibinfo {volume} {33}},\
  \bibinfo {pages} {1830027} (\bibinfo {year} {2018})},\ \Eprint
  {http://arxiv.org/abs/1808.00480} {arXiv:1808.00480 [hep-ph]} \BibitemShut
  {NoStop}%
\bibitem [{\citenamefont {Araz}\ \emph {et~al.}(2020)\citenamefont {Araz},
  \citenamefont {Fuks},\ and\ \citenamefont {Polykratis}}]{Araz:2020lnp}%
  \BibitemOpen
  \bibfield  {author} {\bibinfo {author} {\bibfnamefont {J.~Y.}\ \bibnamefont
  {Araz}}, \bibinfo {author} {\bibfnamefont {B.}~\bibnamefont {Fuks}}, \ and\
  \bibinfo {author} {\bibfnamefont {G.}~\bibnamefont {Polykratis}},\
  }\href@noop {} {\  (\bibinfo {year} {2020})},\ \Eprint
  {http://arxiv.org/abs/2006.09387} {arXiv:2006.09387 [hep-ph]} \BibitemShut
  {NoStop}%
\bibitem [{\citenamefont {Klamke}\ and\ \citenamefont
  {Zeppenfeld}(2007)}]{Klamke:2007cu}%
  \BibitemOpen
  \bibfield  {author} {\bibinfo {author} {\bibfnamefont {G.}~\bibnamefont
  {Klamke}}\ and\ \bibinfo {author} {\bibfnamefont {D.}~\bibnamefont
  {Zeppenfeld}},\ }\href {\doibase 10.1088/1126-6708/2007/04/052} {\bibfield
  {journal} {\bibinfo  {journal} {JHEP}\ }\textbf {\bibinfo {volume} {04}},\
  \bibinfo {pages} {052} (\bibinfo {year} {2007})},\ \Eprint
  {http://arxiv.org/abs/hep-ph/0703202} {arXiv:hep-ph/0703202} \BibitemShut
  {NoStop}%
\bibitem [{\citenamefont {Hagiwara}\ and\ \citenamefont
  {Mukhopadhyay}(2013)}]{Hagiwara:2013jp}%
  \BibitemOpen
  \bibfield  {author} {\bibinfo {author} {\bibfnamefont {K.}~\bibnamefont
  {Hagiwara}}\ and\ \bibinfo {author} {\bibfnamefont {S.}~\bibnamefont
  {Mukhopadhyay}},\ }\href {\doibase 10.1007/JHEP05(2013)019} {\bibfield
  {journal} {\bibinfo  {journal} {JHEP}\ }\textbf {\bibinfo {volume} {05}},\
  \bibinfo {pages} {019} (\bibinfo {year} {2013})},\ \Eprint
  {http://arxiv.org/abs/1302.0960} {arXiv:1302.0960 [hep-ph]} \BibitemShut
  {NoStop}%
\bibitem [{\citenamefont {Nakamura}(2016)}]{Nakamura:2015uca}%
  \BibitemOpen
  \bibfield  {author} {\bibinfo {author} {\bibfnamefont {J.}~\bibnamefont
  {Nakamura}},\ }\href {\doibase 10.1007/JHEP05(2016)157} {\bibfield  {journal}
  {\bibinfo  {journal} {JHEP}\ }\textbf {\bibinfo {volume} {05}},\ \bibinfo
  {pages} {157} (\bibinfo {year} {2016})},\ \Eprint
  {http://arxiv.org/abs/1509.04164} {arXiv:1509.04164 [hep-ph]} \BibitemShut
  {NoStop}%
\end{thebibliography}%

\end{document}